\documentclass[aps,prd,email,preprint,showpacs,showkeys,preprintnumbers,amsmath,amssymb,nofootinbib]{revtex4-2}

\usepackage{color}
\usepackage{latexsym}
\usepackage{amsmath}
\usepackage{amssymb}
\usepackage{eufrak}
\usepackage{euscript}
\usepackage{graphics}
\usepackage{graphicx}
\usepackage{picture}
\usepackage{pstricks,pst-coil}
\usepackage{pst-all}
\usepackage{blindtext}

\newcommand{\be}{\begin{equation}}
\newcommand{\ee}{\end{equation}}
\newcommand{\bn}{\begin{eqnarray}}
\newcommand{\en}{\end{eqnarray}}

\def\[{\left\lbrack}
\def\]{\right\rbrack}

\def\({\left(}
\def\){\right)}
	
\def\MyItem[#1]#2{\item[{#1}]#2}

\def\[{\left\lbrack}
\def\]{\right\rbrack}

\def\({\left(}
\def\){\right)}

\begin{document}


\title{Remarks on classical pseudo-electrodynamics}

\author{S. Duque Cesar$^{a}$}
\email{samduquec@gmail.com}
\author{M. J. Neves$^{b}$}
\email{mariojr@ufrrj.br}

\affiliation{$^a$$^b$Departamento de F\'{i}sica, Universidade Federal Rural do Rio de Janeiro, BR 465-07, 23890-971, Serop\'edica, RJ, Brazil}


\begin{abstract}

Classical studies as the conservation laws and the radiation fields are investigated in the pseudo-electrodynamics. We explore the action symmetry under infinitesimal transformations to obtain the energy-momentum, the Belinfante-Rosenfeld, and the general angular momentum tensors for this nonlocal planar electrodynamics. Through the results such as the retarded potentials and fields generated by a point particle in an arbitrary motion, we study the radiation of an electric dipole and it radiated power in 1+2 dimensions. In addition, we propose a way to introduce magnetic monopoles in pseudo-electrodynamics, in which the solutions and conservation laws are also presented.
\end{abstract}


\maketitle


\section{Introduction}
The connection between condensed matter field and field theories has gained relevance 
in the last decades due to applications in bidimensional materials. The quantum Hall effect \cite{Ando,Tsui,Laughlin,Chamon}, 
planar materials \cite{Qi,Hasan,Chiu,Zhao,Qi2008}, transport in graphene \cite{Gorbar1,Herbut,Gusynin,Herbut2}, 
and superconductivity \cite{Tesanovic,Zhang,Franz,Kivelson,Marino2018} are the examples in which the quantum field theories 
in two dimensions show a successful description of the quantum properties in material physics. The interaction between fermions 
in planar materials is the main motivation to understand the mediator field theory. Based on principles of the quantum electrodynamics (QED), 
the structure is a gauge field theory in $1+2$ dimensions whose the potential links the interaction electron-electron. This gauge theory is known 
in the literature as pseudo-electrodynamics (PED) \cite{Marino93}.       

\pagebreak

The PED is a non-local electromagnetism that arises from the dimensional reduction of the Maxwell's electrodynamics when the classical sources 
are constrained to a two-dimensional space. As consequence, the theory dimensionally reduced is non-local with derivatives of infinity order 
in the D'Alembertian operator. As fundamental properties, the PED preserves the gauge invariance, the theory respects the causality principle \cite{Amaral}, 
and from the point of view of a quantum field theory, the unitarity is also confirmed \cite{Marino2014}. Some extensions of the PED has been studied 
in the literature in last times, as the addition of Chern-Simons topological term \cite{Alves,Alves2}, 
the dimensional reduction in Proca electrodynamics \cite{Ozela,Ozela2,VanSergio}, and the Lee-Wick PED, if the interest it is studying 
a massive pseudo-gauge field with gauge invariance \cite{MarioPRD2025}. The introduction of magnetic monopoles in PED 
also is motivated in two-dimensional topological insulators and in superconductors \cite{XiPRB}. The formulation of two-dimensional magnetic monopole 
gas is another way to investigate the emergence of magnetic monopoles in semiconductor heterostructures \cite{Miao}. 
%


%
In this work, we show some classical results of the pseudo-electrodynamics. We start with a review on the PED.
From the action of the PED, we calculate the energy-momentum tensor and the conserved current via Noether theorem. Since 
the energy-momentum tensor is not gauge invariant, and it is not a symmetric tensor, we construct the approach to find 
the Belinfante-Rosenfeld tensor, that yields the new energy momentum tensor, symmetric and gauge invariant. Thereby, 
we obtain the energy, the linear momentum, the general angular momentum, and the Lorentz force law 
that provides the stress tensor stored in the fields of the PED. Using the retarded solutions for the potentials and Jefimenko's equations, 
we apply it to the case of a time-varying electric dipole placed in a two-dimensional plane in the approximation of the radiation zone. 
Thus, we calculate the power radiated in relation to radial distance. Posteriorly, we present a way to introduce magnetic charge and current distributions in the PED.

The organization of the paper is as follows : In section \ref{sec2} we show a review on pseudo-electrodynamics (PED). The section \ref{sec3} is dedicated to conservation laws and the Noether theorem in the PED. In section \ref{sec4}, we discuss the classical radiation for systems such as electric dipole momentum that depends on time. The section \ref{sec5} presents the pseudo-electrodynamics in the presence of magnetic charge and current distributions. Finally, section \ref{sec6}
is dedicated to the conclusions and perspectives.

We adopt the natural units of $\hbar=c=1$, and the Minkowski metric in space-time of $1+3$ is $\eta_{\mu\nu}=\mathrm{diag}(+1,-1,-1,-1)$. After the dimensional reduction, the metric used is $\eta_{\bar{\mu}\bar{\nu}}=\mathrm{diag}(+1,-1,-1)$, with the greek bar index running as $\bar\mu,\bar\nu=\{ \, 0,1,2 \, \}$.

\section{A review of pseudo-electrodynamics}
\label{sec2}

The well-known classical electrodynamics is governed by the action
\begin{eqnarray}\label{Acaoxi}
S(A^{\mu})=\int_{{\mathcal M}}d^4x \, \left[ \, -\frac{1}{4}\,F_{\mu\nu}F^{\mu\nu}-\frac{1}{2\xi}\,(\partial_{\mu}A^{\mu})^2-J_{\mu}\,A^{\mu} \, \right] \; ,
\end{eqnarray}
where ${\cal M}$ symbolizes the Minkowski spacetime $1+3$, $F_{\mu\nu}=\partial_{\mu}A_{\nu}-\partial_{\nu}A_{\mu}=(-E^{i},-\epsilon^{ijk}\,B^{k})$ is the electromagnetic field tensor, $A^{\mu}$ is the four-potential, $J^{\mu}$ are the classical sources, and $\xi$ is a gauge fixing parameter. As usual, the action principle yields the field equation for the $A^{\mu}$-potential
%
%
%
%
%
\begin{eqnarray}\label{OAJ}
\mathcal{O}_{\mu\nu}\,A^{\mu}(x)=J_{\nu}(x) \; ,
\end{eqnarray}
where $\mathcal{O}_{\mu\nu}:=\eta_{\mu\nu}\,\Box+(\xi^{-1}-1)\,\partial_{\mu}\partial_{\nu}$. 
The solution of (\ref{OAJ}) is
\begin{eqnarray}\label{solAmu}
A_{\mu}(x)=A_{\mu}^{(0)}(x)+\int d^4x' \, \Delta_{\mu\nu}(x-x')\,J^{\nu}(x') \; ,
\end{eqnarray}
in which $A_{\mu}^{(0)}$ is the homogeneous solution of (\ref{OAJ}), and 
$\Delta_{\mu\nu}(x-x')$ is the Green function for the operator $\mathcal{O}_{\mu\nu}$, that satisfies the equation
\begin{eqnarray}\label{eqODelta}
\mathcal{O}_{\mu\nu}\,\Delta^{\nu}_{\;\;\,\alpha}(x-x')=\eta_{\mu\alpha}\,\delta^{4}(x-x') \; .    
\end{eqnarray}
Using the Fourier integral, the result for $\Delta_{\mu\nu}(x-x')$ is
%
%
%
%
%
%
%
%
%
%
%
%
\begin{eqnarray}\label{DeltaFGresult}
\Delta_{\mu\nu}(x-x')=-\int \frac{d^4k}{(2\pi)^4} \, \frac{1}{k^2} \left[\,\eta_{\mu\nu}+(\xi-1) \, \frac{k_{\mu}\,k_{\nu}}{k^2}\, \right] \, e^{-ik\cdot(x-x')} \; .
\end{eqnarray}
%

After integration by parts, the action (\ref{Acaoxi}) is written in the form field-operator-field, namely,
\begin{eqnarray}\label{AcaoxiAOA}
S(A^{\mu})=\int_{{\mathcal M}} d^4x \, \left[ \, \frac{1}{2}\, A_{\mu}\,\mathcal{O}_{\mu\nu} \, A^{\nu}-J_{\mu}\,A^{\mu} \, \right] \; ,
\end{eqnarray}
and substituting (\ref{solAmu}), the action is now a functional of $J^{\mu}$
\begin{eqnarray}\label{AcaoJ}
S(J^{\mu})=-\frac{1}{2}\int_{{\mathcal M}}d^4x \, d^{4}x' \, J_{\mu}(x) \, \Delta^{\mu\nu}(x-x^{\prime}) \, J_{\nu}(x') \; .
\end{eqnarray}
The dimensional reduction is so introduced confining the sources on a spatial plane $2D$ that are constrained by the conditions
\begin{subequations}\label{correnteJ}
\begin{eqnarray}
J^{\bar{\mu}}(x^{\mu}) &=& j^{\bar{\mu}}(x^{\bar{\mu}})\,\delta(z) \; ,
\label{Jmu}
\\
J^{3}(x^{\mu}) &=& 0 \; ,
\label{J3}
\end{eqnarray}
\end{subequations}
where $x^{\bar{\mu}}=(t,x,y)$ are the coordinates defined in the $1+2$ space-time, in which the bar index runs as $\bar{\mu}=\left\{0,1,2\right\}$. Substituting the current (\ref{correnteJ}) in the (\ref{AcaoJ}), and after the ${\cal Z}$-integration, the action is reduced to :
\begin{eqnarray}\label{acao2+1}
\left. S(j^{\bar\mu})=-\frac{1}{2} \int d^3x \, d^3x' \, j_{\bar{\mu}}(\bar{x}) \, \Delta^{\bar{\mu}\bar{\nu}}(x-x^{\prime})\right|_{z=z'=0} \, j_{\bar{\nu}}(\bar{x}') \; ,
\end{eqnarray}
in which the Green function in $1+2$ dimensions is read as
\begin{equation}
\left. \Delta_{\bar{\mu}\bar{\nu}}(x-x^{\prime})\right|_{z=z'=0}=\eta_{\bar{\mu}\bar{\nu}}
\int \frac{d^3k}{(2\pi)^3} \; e^{-i\bar{k}\cdot(\bar{x}-\bar{x}')}\,\int_{-\infty}^{\infty}\frac{dk_z}{2\pi}\frac{1}{k_z^2-\bar{k}^2} \; .
\end{equation}
We have discarded the gauge term of $k_{\bar{\mu}}\,k_{\bar{\nu}}$ by the condition of the conserved current $k_{\bar\mu}\,j^{\bar\mu}=0$ in the momentum space when the Green function is inserted in the action (\ref{acao2+1}). Here, $j^{\bar\mu}=(\sigma,K^i)$ sets a superficial current density. Since $\bar{k}^2=k_0^2-k_x^2-k_y^2 > 0$, or $\bar{k}^2<0$, it is convenient to introduce the coordinates in the Euclidean space-time, where $x_{0}=-i\,x_4$ and $k_0=i\,k_{4}$, such that $\bar{k}^2=-(k_4^2+k_x^2+k_y^2)=-\bar{k}_{E}^2<0$. Thereby, the Green function in Euclidean (and reduced) space-time is
\begin{equation}
\left. \Delta_{E\bar{\mu}\bar{\nu}}(x_{E}-x_{E}^{\prime})\right|_{z=z'=0}=\delta_{\bar{\mu}\bar{\nu}}
\int \frac{d^3\bar{k}_{E}}{(2\pi)^3} \; e^{-i\bar{k}_{E}\cdot(\bar{x}_{E}-\bar{x}_{E}^{\prime})}\,\int_{-\infty}^{\infty}\frac{dk_z}{2\pi}\frac{1}{k_z^2+\bar{k}_{E}^2} \; ,
\end{equation}
the $k_{z}$-integration yields the result
\begin{equation}\label{Delta2+1D}
\Delta_{E\bar{\mu}\bar{\nu}}(\bar{x}-\bar{x}^{\prime})=
-\delta_{\bar{\mu}\bar{\nu}}\int \frac{d^3\bar{k}_E}{(2\pi)^3} \; \frac{e^{-i\bar{k}_{E}\cdot(\bar{x}_{E}-\bar{x}_{E}^{\prime})}}{2\,\sqrt{\bar{k}_{E}^2}} \; .
\end{equation}
%
%
%
Following the inverse way, the task now is to obtain the Lagrangian that leads to correspondent Green function (\ref{Delta2+1D}) in the Euclidean space $3D$. The known solution in the literature \cite{Marino} is the action of the PED in the presence of a non-local gauge fixing term, namely,
%
%
%
\begin{eqnarray}
S_{PED}(A^{\bar{\mu}}) = \int_{\Omega} d^3\bar{x} \; \left[ \, -\frac{1}{4} \bigg(F_{\bar\mu\bar\nu} \frac{2}{\sqrt{\bar{\Box} }}F^{\bar\mu\bar\nu}\bigg)-\frac{1}{2\xi}(\partial_{\bar{\mu}}A^{\bar{\mu}})
\frac{2}{\sqrt{\bar{\Box}}}(\partial_{\bar{\nu}}A^{\bar{\nu}})
-j_{\bar\mu}\, A^{\bar\mu} \, \right] \; ,
\end{eqnarray}
in which $\Omega$ is the Minkowski space of $1+2$ dimensions. In the limit $\xi \rightarrow \infty$, the gauge fixing term is removed, and the Lagrangian density for the PED is 
%
%
\begin{eqnarray}\label{LPED}
{\mathcal L}_{PED}=-\frac{1}{4} \; F_{\bar\mu\bar\nu}\,\bigg[ \, \frac{2}{\sqrt{\bar{\Box}}}F^{\bar\mu\bar\nu} \, \bigg]-j_{\bar\mu}\,A^{\bar\mu} \; ,
\end{eqnarray}
where $F_{\bar\mu\bar\nu}=\partial_{\bar\mu}A_{\bar\nu}-\partial_{\bar\nu}A_{\bar\mu}=(-E^{i},-B_{z}) \, (i=1,2)$ is the EM strength field tensor, $A^{\bar\mu}=(A^{0},A^{i})$ is the $3$-potential, the operator with bar index means the derivatives in relation to $(t,x,y)$, $\partial_{\bar\mu}=(\partial_{t},\partial_{x},\partial_{y})$, and $\bar{\Box}=\partial_{\bar\mu}\partial^{\bar\mu}=\partial_{t}^2-\partial_{x}^2-\partial_{y}^2$ is the D'Alembertian operator in $1+2$ dimensions. The non-local operator $2/\sqrt{\bar{\Box}}$ in the kinetic term can be written as the infinity series
\begin{eqnarray}\label{Oplimit}
\frac{2}{\sqrt{\bar{\Box}}}=
\lim_{\alpha \rightarrow 0} \frac{2}{\alpha}\frac{1}{\sqrt{1+\frac{\bar{\Box}}{\alpha^2} }}=\lim_{\alpha \rightarrow 0} \; \sum_{n=0}^{\infty} 
\, \frac{ 2\,\sqrt{\pi} \,\alpha^{-2n-1}}{ n! \, \Gamma(1/2-n)} \, \bar{\Box}^n \; ,   
\end{eqnarray}
in which $\bar{\Box}^n=\prod_{i=1}^{n}\bar{\Box}$, if $n > 0$, and $\Box^n=1$, if $n=0$. After some integrations by parts in the action, the Lagrangian is written in the form 
\begin{eqnarray}\label{LPEDOp}
{\mathcal L}_{PED}=\frac{1}{2} \; A_{\bar\mu}\,\bigg[ \, 
(\eta^{\bar\mu\bar\nu}\,\bar{\Box}-\partial^{\bar\mu}\,\partial^{\bar\nu})\lim_{\alpha \rightarrow 0} \; \sum_{n=0}^{\infty} 
\, \frac{ 2\,\sqrt{\pi} \,\alpha^{-2n-1}}{ n! \, \Gamma(1/2-n)} \, \bar{\Box}^n  \, \bigg]A_{\bar\nu}-j_{\bar\mu}\;A^{\bar\mu} \; ,
\end{eqnarray}
and shows that the PED is a field theory with derivatives of infinite order in the D'Alembertian operator $\bar{\Box}$. The action correspondent to the lagrangian density (\ref{LPED}) is invariant under the gauge transformation $A_{\bar\mu} \, \longmapsto \, A_{\bar{\mu}}^{\prime}=A_{\bar\mu}+\partial_{\bar\mu}f$, for any $f$-scalar function, if the current density is conserved, {\it i.e.}, 
$\partial_{\bar{\mu}}j^{\bar{\mu}}=0$.

Using the action principle, the field equations for the classical pseudo-electrodynamics in the presence of $j^{\bar{\mu}}$ are
\begin{equation}\label{Eqcampoj}
\frac{2}{\sqrt{\Box }}(\partial^{\bar\mu}F_{\bar\mu\bar\nu}) = j_{\bar\nu} \; ,
\end{equation}
and the Bianchi identity keeps like in the usual ED, but with the bar index in $1+2$ dimensions
\begin{eqnarray}\label{IdBianchi}
\partial^{\bar\mu}F^{\bar\nu\bar\rho}+ \partial^{\bar\nu}F^{\bar\rho\bar\mu}+\partial^{\bar\rho}F^{\bar\mu\bar\nu}=0 \; .    
\end{eqnarray}
Alternatively, the Bianchi identity also can be represented by $\partial_{\bar{\mu}}\tilde{F}^{\bar{\mu}}=0$, in which $\tilde{F}_{\bar{\mu}}=\epsilon_{\bar\mu\bar\nu\bar\rho}\,F^{\bar\nu\bar\rho}/2$ is the dual tensor of $F_{\bar\mu\bar\nu}$, in which it has the following components $\tilde{F}^{\bar{\mu}}=(B_{z},E^{i})$.

In vector form, the equations in terms of the components 
$E^{i} \, (i=1,2)$ and $B_{z}$ are given by
%
\begin{eqnarray}
\frac{2}{\sqrt{\bar{\Box}}}\left[ \, \bar{\nabla}\cdot{\bf E}  \, \right]=\sigma
\hspace{0.3cm} , \hspace{0.3cm}
\frac{2}{\sqrt{\bar{\Box}}}\left[ \, \epsilon^{ij}\,\partial_{j}B_{z}-\partial_{t}E^{i} \, \right]=K^{i}
\hspace{0.3cm} , \hspace{0.3cm}
(\bar{\nabla}\times{\bf E})_{z}+\partial_{t}B_{z}=0 
\; ,
\end{eqnarray}
%
where the spatial operator $\bar{\nabla}=(\partial_{x},\partial_{y})$ means the derivatives in relation to coordinates $x$ and $y$. Due the gauge symmetry mentioned previously, we choose the non-local gauge $2/\sqrt{\bar{\Box}}(\partial_{\bar{\mu}}A^{\bar{\mu}})=0$, such that the equation for the $3$-potential $A^{\bar{\mu}}$ is
\begin{eqnarray}\label{EqA}
2\,\sqrt{\bar{\Box}}\, A^{\bar{\mu}}(\bar{x}) = j^{\bar{\mu}}(\bar{x}) \; .
\end{eqnarray}
The solution of (\ref{EqA}) is obtained by the Green function method via Fourier transform, as in the usual case. Since we are interested in causal propagation and in the radiation, the retarded Green function is the same from Maxwell ED \cite{Marino}, in which the result is
\begin{eqnarray}
A^{\bar\mu}(\bar{{\bf r}},t)=\frac{1}{4\pi}\int_{\mathcal{S}} d^2\bar{{\bf r}}^{\prime} 
\, \, \frac{j^{\bar{\mu}}(\bar{{\bf r}}^{\prime},t_r)}{|\bar{{\bf r}}-\bar{{\bf r}}^{\prime}|} \; ,
\end{eqnarray}
where $t_r=t-|\bar{{\bf r}}-\bar{{\bf r}}^{\prime}|$ is the retarded time, and the integral is over the spatial plane $\mathcal{S}$. Thereby, the non-local operator of this theory does not change the interpretation of the retarded propagation for any source of charges and currents on the spatial plane, and as consequence, the Green function preserves the causality in the PED \cite{Marino}.


\section{Noether theorem and the Belinfante-Rosenfeld tensor in PED}
\label{sec3}

The concepts of symmetries and conserved quantities are linked by the Noether theorem. The invariance of the action under infinitesimal transformations of the space-time coordinates and of fields leads to physical conserved quantities.  
%
%
We start with the infinitesimal transformations in the coordinates, 
and in the $3$-potential of the PED  
%
%
\begin{subequations}
\begin{eqnarray}
x^{\bar{\mu}} \; \mapsto \; x^{\bar{\mu}} &=& x^{\bar{\mu}}+\delta x^{\bar{\mu}} \; ,
\label{transfx}
\\
A^{\bar{\mu}}(x) \; \mapsto \; A^{'\bar{\mu}}(x') &=& A^{\bar{\mu}}(x)+\delta A^{\bar{\mu}}(x) \; ,
\label{transfA}
\end{eqnarray}
\end{subequations}
where $\delta x^{\bar{\mu}}$ is a small variation of the coordinates $x^{\bar\mu}$. The transformations (\ref{transfx}) and (\ref{transfA}) 
lead to variation in the lagrangian of the PED as,  $\mathcal{L}_{PED} \, \longmapsto \, \mathcal{L}'_{PED}=\mathcal{L}_{PED}+\delta \mathcal{L}_{PED}$, 
and the action variation is
\begin{eqnarray}\label{deltaSPEDNoether}
\delta S_{PED}
=\int_{\Omega^{\prime}} d^3x' \; 
\mathcal{L}_{PED}(A'_{\bar{\nu}}(x'),\bar{\Box^{\prime}}\,A'_{\bar{\nu}}(x'),\ldots,\bar{\Box^{\prime}}^n\,A'_{\bar{\nu}}(x'))
\nonumber \\
-\int_{\Omega} d^3x \; \mathcal{L}_{PED}(A_{\bar{\nu}}(x),\bar{\Box} \, A_{\bar{\nu}}(x),\ldots,\bar{\Box}^n\,A_{\bar{\nu}}(x)) \; .  
\end{eqnarray}
The variation of $A^{\bar{\mu}}$ in the same point $x^{\bar\mu}$ is defined by
%
$\tilde{\delta}A^{\bar\nu}=A^{'\bar\nu}(x)-A^{\bar\nu}(x)$,
%
that can be written as
%
%
%
$\tilde{\delta}A^{\bar\nu}=\delta A^{\bar\nu}-(\partial_{\bar\mu}A^{\bar\nu})\;\delta x^{\bar\mu}$.  
%
Thereby, the variation of the Lagrangian in $x^{\bar\mu}$ is
%
$\tilde{\delta}\mathcal{L}_{PED}=\delta \mathcal{L}_{PED}-(\partial_{\bar\mu}\mathcal{L}_{PED})\,\delta x^{\bar\mu}$.
%
Using these results in (\ref{deltaSPEDNoether}), the variation of the action is 
\begin{eqnarray}
\delta S_{PED}=\int_{\Omega} d^3x \; \left[ \, \tilde{\delta}\mathcal{L}_{PED}+\partial_{\bar\mu}(\delta x^{\bar\mu} \, \mathcal{L}_{PED}) \, \right] \; .
\end{eqnarray}
The equation (\ref{Eqcampoj}), without source $j^{\bar\mu}=0$, yields the variation of the Lagrangian in $x^{\bar\mu}$ 
\begin{eqnarray}
\tilde{\delta}\mathcal{L}_{PED}=-\partial_{\bar\mu}\left[ \, \left( \frac{2}{\sqrt{\bar{\Box}}}F^{\bar\mu\bar\nu}\right)\tilde{\delta}A_{\bar\nu}\, \right] \; ,
\end{eqnarray}
thus, the variation of the action is    
\begin{eqnarray}\label{deltaSPEDf}
\delta S_{PED}=\int_{\Omega} d^3x \; \partial_{\bar\mu}f^{\bar\mu} \; ,    
\end{eqnarray}
where $f^{\bar\mu}$ is the Noether current
\begin{eqnarray}\label{Noethercurrent}
f^{\bar\mu}=-\left[\frac{2}{\sqrt{\bar{\Box}}}F^{\bar\mu\bar\nu} \right]\delta A_{\bar\nu}-\delta x_{\bar\alpha}\;\Theta^{\bar\mu\bar\alpha} \; ,    
\end{eqnarray}
and $\Theta^{\bar\mu\bar\alpha}$ is the energy-momentum tensor
\begin{eqnarray}\label{tensorEM}
\Theta^{\bar\mu\bar\alpha}=-\left[\frac{2}{\sqrt{\bar{\Box}}}F^{\bar\mu\bar\nu}\right]\,\partial^{\bar\alpha}A_{\bar\nu}-\eta^{\bar\mu\bar\alpha} \; \mathcal{L}_{PED} \; .
\end{eqnarray}
The invariance of the action under (\ref{transfx}) and (\ref{transfA}) is such that, $\delta S_{PED}=0$, and the result (\ref{deltaSPEDf}) yields the continuity equation 
%
$\partial_{\bar\mu}\,f^{\bar\mu}=0$, 
%
%
%
%
that when integrated over all spatial plane $(\mathcal{S})$ leads to Noether's conserved charge
\begin{eqnarray}
Q^{(N)}=\int_{\mathcal{S}} d^2\bar{x} \; f^{0} \; .    
\end{eqnarray}
%
%

%
As we can see in (\ref{tensorEM}), the energy-momentum tensor is not gauge invariant, and also it is not a symmetric tensor. Analogous to the usual electromagnetism, these 
properties must be recovered introducing the Belifante-Rosenfeld tensor. This approach requires the definition of the angular momentum for the $A^{\bar\mu}$ gauge field in the 
context of the pseudo-electrodynamics. Under the Lorentz group in $1+2$ dimensions, the infinitesimal variations of $x^{\bar\mu}$ and $A^{\bar\mu}$ are, respectively, given by    
\begin{subequations}
\begin{eqnarray}
\delta x^{\bar\alpha} &=& \omega^{\bar\alpha\bar\beta} \, x_{\bar\beta} \; ,
\\    
\delta A_{\bar\nu} &=& \frac{1}{2} \, \omega_{\bar\alpha\bar\beta} 
\, (-i\,\Sigma^{\bar\alpha\bar\beta})_{\bar\nu\bar\sigma} \, A^{\bar\sigma} \; ,
\end{eqnarray}    
\end{subequations}
where $\omega^{\bar\alpha\bar\beta}$ are real parameters, in which 
$\omega^{\bar\alpha\bar\beta}=-\omega^{\bar\beta\bar\alpha}$, and $(\Sigma^{\bar\alpha\bar\beta})_{\bar\nu\bar\sigma}=i\left(\delta^{\bar\alpha}_{\;\;\,\bar\sigma}
\,\delta^{\bar\beta}_{\;\;\,\bar\nu}-\delta^{\bar\alpha}_{\;\;\,\bar\nu}
\,\delta^{\bar\beta}_{\;\;\,\bar\sigma}\right)$ is the generator for the vector representation in the Lorentz group. Substituting these variations in (\ref{Noethercurrent}), the Noether current is written as
\begin{eqnarray}
f^{\bar\mu}=\frac{1}{2} \, \omega_{\bar\nu\bar\alpha} \, M^{\bar\mu\bar\nu\bar\alpha} \; ,    
\end{eqnarray}
in which the tensor $M^{\bar\mu\bar\nu\bar\alpha}$ is defined by
\begin{eqnarray}
M^{\bar\mu\bar\nu\bar\alpha}=\Theta^{\bar\mu\bar\alpha} \, x^{\bar\nu} 
- \Theta^{\bar\mu\bar\nu} \, x^{\bar\alpha}
+\left[\frac{2}{\sqrt{\bar{\Box}}}F^{\bar\mu\bar\nu} \right]\,A^{\bar\alpha}
-\left[\frac{2}{\sqrt{\bar{\Box}}}F^{\bar\mu\bar\alpha} \right]\,A^{\bar\nu} \; .    
\end{eqnarray}
For $\omega^{\bar\alpha\bar\beta} \neq 0$, the continuity equation implies that $\partial_{\bar\mu}M^{\bar\mu\bar\nu\bar\alpha}=0$, and the conserved angular momentum is
\begin{eqnarray}
N^{\bar\nu\bar\alpha}=\int_{\mathcal{S}}d^2x \; M^{0\bar\nu\bar\alpha}  \; ,  
\end{eqnarray}
in which $N^{\bar\nu\bar\alpha}=-N^{\bar\alpha\bar\nu}$ is antisymmetric, and therefore, 
it has three components $N^{\bar\nu\bar\alpha}=(N^{01},N^{02},N^{12})$ in $1+2$ dimensions. The spatial component is defined as the general angular momentum on ${\cal Z}$-direction, {\it i.e.}, $N^{12}=J_{z}=L_{z}+S_{z}$, where the orbital angular momentum $(L_{z})$ and the spin $(S_{z})$ are given by
\begin{subequations}
\begin{eqnarray}
L_{z} \!&=&\! \int_{\mathcal{S}} d^2\bar{x} \, \left\{ \left[ \frac{2}{\sqrt{\bar{\Box}}} {\bf E} \right] \cdot(x\,\partial_y-y\,\partial_x){\bf A} \right\} \; ,
\\
S_{z} \!&=&\! \int_{\mathcal{S}} d^2\bar{x} \, \left\{ \, \left[ \frac{2}{\sqrt{\bar{\Box}}} E_{x} \right] \, A_{y}-\left[ \frac{2}{\sqrt{\bar{\Box}}} E_{y} \right] \, A_{x} \, \right\} \; ,
\end{eqnarray}
\end{subequations}
respectively.
The new energy-momentum tensor (symmetric and gauge invariant) is so written as $T^{\bar\mu\bar\nu}=\Theta^{\bar\mu\bar\nu}+\partial_{\bar\lambda}X^{\bar\lambda\bar\mu\bar\nu}$, where $X^{\bar\lambda\bar\mu\bar\nu}=-X^{\bar\mu\bar\lambda\bar\nu}$, and $T^{\bar\mu\bar\nu}$ also satisfies the continuity equation $\partial_{\bar\mu}T^{\bar\mu\bar\nu}=0$. Thereby, the modified angular momentum tensor is $\tilde{M}_{\bar\mu\bar\nu\bar\lambda}=T_{\bar\mu\bar\lambda}\,x_{\bar\nu}-T_{\bar\mu\bar\nu}\,x_{\bar\lambda}$, where $\partial^{\bar\mu}\tilde{M}_{\bar\mu\bar\nu\bar\lambda}=0$, if the energy-momentum tensor is symmetric, {\it i.e.}, $T_{\bar\mu\bar\nu}=T_{\bar\nu\bar\mu}$. Writing $\tilde{M}_{\bar\mu\bar\nu\bar\lambda}=M_{\bar\mu\bar\nu\bar\lambda}+\partial^{\bar\sigma}Y_{\bar\sigma\bar\mu\bar\nu\bar\lambda}$, the $Y_{\bar\sigma\bar\mu\bar\nu\bar\lambda}$ tensor satisfies the condition $Y_{\bar\sigma\bar\mu\bar\nu\bar\lambda}=-Y_{\bar\mu\bar\sigma\bar\nu\bar\lambda}$, and substituting the definitions of $M_{\bar\mu\bar\nu\bar\rho}$ and $\tilde{M}_{\bar\mu\bar\nu\bar\rho}$, we obtain the relation
\begin{eqnarray}
(\partial^{\bar\beta}X_{\bar\beta\bar\mu\bar\lambda})\,x_{\bar\nu}
-(\partial^{\bar\beta}X_{\bar\beta\bar\mu\bar\nu})\,x_{\bar\lambda}= \left[ \frac{2}{\sqrt{\bar{\Box}}}F_{\bar\mu\bar\nu} \right]\,A_{\bar\lambda}-\left[\frac{2}{\sqrt{\bar{\Box}}}F_{\bar\mu\bar\lambda}\right]\,A_{\bar\nu}+\partial^{\bar\sigma}Y_{\bar\sigma\bar\mu\bar\nu\bar\lambda} \; .   
\end{eqnarray}
The solution for the tensor $X_{\bar\nu\bar\mu\bar\lambda}$ is
%
\begin{eqnarray}
X_{\bar\mu\bar\nu\bar\lambda}=-\left[ \, \frac{2}{\sqrt{\bar{\Box}}}F_{\bar\mu\bar\nu} \, \right]\,A_{\bar\lambda} \; ,
\end{eqnarray}
that satisfies the properties of permutation of the index, shown previously, in which $Y_{\bar\sigma\bar\mu\bar\nu\bar\lambda}=x_{\bar\nu}\,X_{\bar\sigma\bar\mu\bar\lambda}-x_{\bar\lambda}\,X_{\bar\sigma\bar\mu\bar\nu}$. 
Therefore, the symmetric energy-momentum tensor is
\begin{eqnarray}
T^{\bar\mu\bar\nu} = \bigg[\,\frac{2}{\sqrt{\bar{\Box}}}F^{\bar\mu\bar\alpha}\,\bigg]F^{\bar\nu}_{\;\;\;\;\bar\alpha} + \frac{1}{4} \;\eta^{\bar\mu\bar\nu} \; \bigg[ \, \frac{2}{\sqrt{\bar{\Box}}} F^{\bar\alpha\bar\beta} \, \bigg] F_{\bar\alpha\bar\beta} \; .
\end{eqnarray}
The continuity equation $\partial_{\bar\mu}T^{\bar\mu\bar\nu}=0$ leads to the conserved momentum
\begin{eqnarray}
P^{\bar\nu} = \int_{\mathcal{S}} d^2\bar{x} \; T^{0\bar{\nu}} \; ,
\end{eqnarray}
where the components $U=P^{0}$ and $P^{i}$ yield the energy, 
and the linear momentum of the gauge field in PED, respectively,
\begin{subequations}
\begin{eqnarray}
U \!&=&\! \int_{\mathcal{S}} d^2\bar{x} \; \bigg[ \, \textbf{E} \cdot\bigg(\frac{1}{\sqrt{\bar{\Box}}} \textbf{E} \bigg) + B_{z} \bigg(\frac{1}{\sqrt{\bar{\Box}}} B_{z} \bigg) \, \bigg] \; , 
\\
P^{i} \!&=&\! \int_{\mathcal{S}} d^2\bar{x} \; \varepsilon^{ij} \, \left[ \, \frac{2}{\sqrt{\bar{\Box}}}E^{j} \, \right] B_{z} \; .
\label{Poynting}
\end{eqnarray}
\end{subequations}
Notice that the density of linear momentum is the Poynting vector of the theory.  The conservation of the modified angular momentum $\tilde{M}^{\bar\mu\bar\nu\bar\lambda}$ contains the conserved components 
\begin{eqnarray}
\tilde{N}^{\bar\nu\bar\lambda}=\int_{{\mathcal S}}d^2\bar{x} \, \tilde{M}^{0\bar\nu\bar\lambda} \; , 
\end{eqnarray}
whose the spatial component $J_{z}=\epsilon^{ij}\,\tilde{N}^{ij}/2$ is the general angular momentum on the ${\cal Z}$-direction
\begin{eqnarray}\label{Jz}
J_{z}=-\int_{{\mathcal S}}d^{2}\bar{x} \, \left\{ x\left[ \frac{2}{\sqrt{\bar{\Box}}} E_{x} \right]+y\left[\frac{2}{\sqrt{\bar{\Box}}} E_{y} \right] \right\}B_{z} \; ,
\end{eqnarray}
that is gauge invariant.

Alternatively, the conservation laws for the symmetric tensor $T^{\bar\mu\bar\nu}$ also can be obtained directly from the field equations (\ref{Eqcampoj}) and (\ref{IdBianchi}). Contracting (\ref{Eqcampoj}) with $F_{\bar\nu\bar\rho}$, with help of (\ref{IdBianchi}), we obtain 
\begin{eqnarray}\label{eqTJ}
\partial_{\bar\mu}T^{\bar\mu\bar\nu}=j_{\bar\rho}\,F^{\bar\rho\bar\nu} \; .   
\end{eqnarray}
For $\bar\nu=j$, the expression (\ref{eqTJ}) is reduced to
\begin{eqnarray}
\partial_{i}\Theta^{i}_{\,\,j}-\partial_{t}S_{j}=f_{j} \; ,    
\end{eqnarray}
where $f_{j}=\sigma\,E_{j}+\varepsilon_{ij}\,j^{i}\,B_{z}$ is the Lorentz force density, and $\Theta^{i}_{\,\,j}=-T^{i}_{\,\,j}$ is the stress tensor of the PED
\begin{eqnarray}
\Theta^{i}_{\,\,j}=\left[\frac{2}{\sqrt{\bar{\Box}}}E^{i}\right]E_{j}
-\delta^{i}_{\,\,j} \, {\bf E}\cdot\left[ \, \frac{1}{\sqrt{\bar{\Box}}}{\bf E} \, \right]    
-\delta^{i}_{\,\,j} \, B_{z}\left[ \, \frac{1}{\sqrt{\bar{\Box}}}B_{z} \, \right] \; .
\end{eqnarray}
This result is usefull for the calculus of force and pressure on the surfaces due to external sources, as example. Thereby, we have obtained all the conserved physical quantities that are gauge invariant in the PED. These quantities can yield results of physical interest as the energy and the radiation power that will be investigated in the next section.   

\section{ Classical radiation in pseudo-electrodynamics}
\label{sec4}
The Green's function method shows that the solution for the $A^{\bar\mu}$-potential in PED is similar to the Maxwell ED. Thus, the causality principle is preserved in PED, even if it is a non-local theory \cite{Marino}. The retarded solutions for the components of the potential are read as   
\begin{eqnarray}
\textbf{A}(\textbf{r}, t) = \frac{1}{4\pi}\int_{\mathcal{S}}\frac{\textbf{K}(\textbf{r}', t_r)}{|\textbf{r} - \textbf{r}'|} \; d^2\textbf{r}' \;,\quad  V(\textbf{r}, t)=\frac{1}{4\pi}\int_{\mathcal{S}}\frac{\sigma(\textbf{r}', t_r)}{|\textbf{r} - \textbf{r}'|} \; d^2\textbf{r}' \; ,
\end{eqnarray}
where $t_{r}=t-|\textbf{r}-\textbf{r}^{\prime}|=t-R$ is the retarded time, and ${\bf R}=\textbf{r}-\textbf{r}^{\prime}=(x-x^{\prime})\,\hat{{\bf x}}+(y-y^{\prime})\,\hat{{\bf y}}$ is the position vector of the sources in relation to an arbitrary point on the spatial plane. The Jefimenko's equations are read below
\begin{subequations}
\begin{eqnarray}
\textbf{E} (\textbf{r}, t) &=& \frac{1}{4\pi} \int_{\mathcal{S}}  d^2\textbf{r}^{\prime} \bigg[ \, \frac{\sigma(\textbf{r}', t_r)}{R^2} \, \hat{\textbf{R}} 
+ \frac{\dot\sigma(\textbf{r}', t_r)}{ R} \, \hat{ \textbf{R}}
-\frac{\dot{\textbf{K}}(\textbf{r}', t_r)}{R} \, \bigg] \; ,
\\
B_z(\textbf{r}, t) &=& \frac{1}{4\pi} \int_{\mathcal{S}}  \; d^2\textbf{r}^{\prime} \; \bigg[ \, \frac{\textbf{K}(\textbf{r}', t_r)}{R^2}+\frac{\dot{\textbf{K}}(\textbf{r}', t_r)}{ R} \, \bigg]\times \left. \hat{ \textbf{R}} \right|_{z} \; ,
\label{IntBz}
\end{eqnarray}
\end{subequations}
and we take the ${\cal Z}$-component of the vector product in the integral (\ref{IntBz}).

%

%
To simplify our analysis and the examples that we are interested, we introduce the approximation of the radiation zone, in which 
$r \gg r^{\prime}$, with $R \simeq r$, such that the electric and magnetic fields of an electric dipole varying with the time 
on the spatial plane are the known expressions in the literature of the Maxwell ED \cite{Jackson} :
\begin{subequations}
\begin{eqnarray}
{\bf E}({\bf r},t) &=& \frac{1}{4\pi\,r} \, \left[ \, \hat{{\bf r}} \times ( \hat{{\bf r}}\times \ddot{{\bf p}}(t_0)) \, \right] \; ,
\\
{\bf B}({\bf r},t) &=& \frac{1}{4\pi\,r}\,\left[ \, \ddot{{\bf p}}(t_0)\times\hat{{\bf r}} \, \right] \; ,
\end{eqnarray}
\end{subequations}
where the retarded time is simplified as $t_0=t-r$, ${\bf r}$ is the radial vector on the plane, and 
$\ddot{{\bf p}}$ is the second derivative of the electric dipole vector ${\bf p}(t_0)$ in relation
to the time $(t)$. As an application, we consider a time-varying electric dipole 
$\textbf{p}=p(t) \, \hat{\textbf{x}}$ along the ${\cal X}$-axis, with the polar 
coordinates $(r,\theta)$, the electric and magnetic fields generated 
by this dipole at a generic point on the ${\cal XY}$ plane are given by :
\begin{eqnarray}
\textbf{E}(r,\theta) = \frac{\ddot{p}(t_0)}{4\pi r} \; \sin\theta \, \hat{\boldsymbol{\theta}}
\quad \mbox{and} \quad
B_{z}(r,\theta) = \frac{\ddot{p}(t_0)}{4\pi r} \; \sin\theta 
\; .
\end{eqnarray}
%
%
Using (\ref{Poynting}), we calculate d'Alembertian operator acting on the electric field, which is as follows
\begin{eqnarray}
\Box \textbf{E}=\bigg[ \frac{\dddot{p}\,(t_0)}{2\pi r}\,\dot{\theta}+\frac{\ddot{p}\,(t_0)}{4\pi r}\,\ddot{\theta} \bigg]\big[ -\sin(2\theta) \, \hat{{\bf x}}+\cos(2\theta) \, \hat{{\bf y}} \big] 
+ \bigg[ -\frac{\ddot{p}(t_0)}{2\pi r}\dot{\theta}^2 + 
\nonumber \\
+\frac{\ddot{p}(t_0)}{2\pi r^3} \bigg]\big[ \cos(2\theta) \, \hat{{\bf x}}+\sin(2\theta) \, \hat{{\bf y}} \big]-\bigg[ \frac{\ddot{p}(t_0)}{4\pi r^3}+\frac{\dddot{p}(t_0)}{4\pi r^2} \bigg]\sin\theta \, \hat{\boldsymbol{\theta}} \; ,
\end{eqnarray}
where the terms in $1/r^2$ and $1/r^3$ do not contribute to the radiation zone. The terms that depends on $(r,\theta)$ are fixed in the plane, and we can adopt $\dot{\theta}=\ddot{\theta}=0$. After some algebraic manipulations, we obtain
\begin{equation}
\Box \textbf{E}= \bigg\{ \frac{\ddot{p}(t_0)}{4\pi r^3}\big[ 
\cos^2\theta+\cos(2\theta) \big]+\frac{\dddot{p}(t_0)}{4\pi r^2}\sin^2\Theta \bigg\}\,\hat{{\bf x}}+\bigg\{ \frac{3\ddot{p}(t_0)}{8\pi r^3} \, \sin(2\theta) - \frac{\dddot{p}\,(t_0)}{8\pi r^2} \, \sin(2\theta) \bigg\} \, \hat{{\bf y}} \; .
\end{equation}
Regarding the components of the Poynting vector 
\begin{subequations}
\begin{eqnarray}
S_x &=& \epsilon^{12}\bigg[ \frac{2}{\sqrt{\bar{\Box}}}E_y \bigg]B_z = \bigg[ \frac{2}{\sqrt{\bar{\Box}}}E_y \bigg]B_z \; ,
\label{Sx}
\\
S_y &=& \epsilon^{21}\bigg[ \frac{2}{\sqrt{\bar{\Box}}}E_x \bigg]B_z = -\bigg[ \frac{2}{\sqrt{\bar{\Box}}}E_x \bigg]B_z \; ,
\label{Sy}
\end{eqnarray}
\end{subequations}
where the non-local operator acts on the electric field components as
\begin{eqnarray}
\frac{2}{\sqrt{\bar{\Box}}}E_{i} = \lim_{\alpha \to 0} \, \sum_{n \to 0}^{\infty} \frac{2\,\sqrt{\pi} \, \alpha^{-2n-1}}{n! \, \Gamma(\frac{1}{2}-n)}\prod_{i=1}^{n}\bar{\Box} \, E_{i} \; ,
\end{eqnarray}
with $i=1,2$. Assuming the angular position at $\theta=\frac{\pi}{4}$ for a radial direction $r$, the components in (\ref{Sx}) and (\ref{Sy}) are, respectively, given by
\begin{subequations}
\begin{eqnarray}
 S_{x} &=& \frac{2\,\ddot{p}(t_0) }{\sqrt{ \, \ddot{p}(t_0)/r+\dddot{p}(t_0) }} \; ,
 \\
 S_{y} &=& \frac{2\,\ddot{p}(t_0) }{\sqrt{ \, 3\ddot{p}(t_0)/r-\dddot{p}(t_0) }} \; .
\end{eqnarray}
\end{subequations}
The power radiated by the electric dipole with constant acceleration 
$\ddot{p}=q\,\ddot{d}=q\,a$ and $\dddot{p}=0$, along the radial direction $(\theta=\pi/4)$ per length unity is
\begin{eqnarray}\label{dPdr}
\frac{dP}{dr} = \mathbf{S} \cdot \hat{\textbf{r}}
=\sqrt{\frac{2}{3}}\,\left(\sqrt{3}-1\right)\,\sqrt{q\,a\,r} \; .
\end{eqnarray}
The variation of the power in relation to radial direction is proportional to squared root of $r$, and also depends on the $\sqrt{a}$, for a positive acceleration. This result is due to the non-locality of the PED, that is completely different from the Larmor's potency known in the usual Maxwell's ED \cite{Jackson}.

%
%


%
\section{Pseudo-electrodynamics with magnetic monopoles}
\label{sec5}

From the PED equations, there is not the divergent of the 
magnetic field in $1+2$ dimensions, and therefore, the 
introduction of a magnetic charge density is not direct like 
in the case of the Maxwell ED in $1+3$ dimensions. 
In the context of the PED, we introduce the new strength field tensor associated with the electric charge as $F_{\bar\mu\bar\nu} \rightarrow F_{e\bar\mu\bar\nu}=\partial_{\bar\mu}A_{e\bar\nu}-\partial_{\bar\nu}A_{e\bar\mu}=(-E_{e}^{\,\,i},B_{ez})$, and also, the strength field tensor associated with the magnetic charge distribution, that is, $F_{m\bar\mu\bar\nu}=\partial_{\bar\mu}A_{m\bar\nu}-\partial_{\bar\nu}A_{m\bar\mu}=(-B_{m}^{\,\,i},E_{mz})$, where $A_{m}^{\,\,\bar\mu}=(\Phi_{m},{\bf A}_m)$ is the magnetic $3$-potential, and $A_{e}^{\,\,\bar\mu}=(\Phi_{e},{\bf A}_e)$ is the electric $3$-potential. Based on the approach for the usual ED in $1+2$ dimensions \cite{McDonald}, the PED in the presence of magnetic charges and currents 
is represented by the set of equations
\begin{subequations}
\begin{eqnarray}
\frac{2}{\sqrt{\bar{\Box}}}\left[\,\partial_{\bar\mu}F_{e}^ {\,\,\bar\mu\bar\nu}\,\right] &=& j_{e}^{\,\,\bar\nu}
\hspace{0.5cm} , \hspace{0.5cm} 
\partial_{\bar\mu}F_{e\bar\nu\bar\rho}
+\partial_{\bar\nu}F_{e\bar\rho\bar\mu}
+\partial_{\bar\rho}F_{e\bar\mu\bar\nu}=0 \; ,
\label{EqsFe}
\\
\frac{2}{\sqrt{\bar{\Box}}}\left[\,\partial_{\bar\mu}F_{m}^ {\,\,\bar\mu\bar\nu}\,\right] &=& j_{m}^{\,\,\bar\nu}
\hspace{0.5cm} , \hspace{0.5cm} 
\partial_{\bar\mu}F_{m\bar\nu\bar\rho}
+\partial_{\bar\nu}F_{m\bar\rho\bar\mu}
+\partial_{\bar\rho}F_{m\bar\mu\bar\nu}=0 \; ,
\label{EqsFm}
\end{eqnarray}
\end{subequations}
in which $j_{m}^{\,\,\bar\nu}=(\sigma_{m},{\bf K}_m)$ is the magnetic $3$-current on the spatial plane, and $j_{e}^{\,\,\bar\nu}=(\sigma_{e},{\bf K}_e)$ is the electric $3$-current. The matrix representation for EM tensors is 
\begin{eqnarray}
F_e^{\,\,\bar\mu\bar\nu}= 
\left[\begin{array}{ccc}
0 & -E_{ex} & -E_{ey} \\
E_{ex} & 0 & -B_{ez} \\
E_{ey} & B_{ez} & 0 \\
\end{array}
\right]
\; \; , \; \;
F_{m}^{\,\,\;\bar\mu\bar\nu}= 
\left[\begin{array}{ccc}
0 & -B_{mx} & -B_{my} \\
B_{mx} & 0 & E_{mz} \\
B_{my} & -E_{mz} & 0 \\
\end{array}
\right] \; ,
\end{eqnarray}
and the field equations in the vector form are read as
\begin{subequations}
\begin{eqnarray}
\frac{2}{\sqrt{\bar{\Box}}}\left[ \, \bar{\nabla}\cdot{\bf E}_{e}  \, \right] &=& \sigma_{e}
\; , \;
\frac{2}{\sqrt{\bar{\Box}}}\left[ \, \epsilon^{ij}\partial_{j} {B}_{ez} -\partial_{t}E_{e}^{\,\,i} \, \right]=K_{e}^{\,\,i} 
\; , \;
(\bar{\nabla}\times{\bf E}_{e})_{z}+\partial_{t}B_{ez}=0 \; ,
\label{eqsE}
\\ 
\frac{2}{\sqrt{\bar{\Box}}}\left[ \, \bar{\nabla}\cdot{\bf B}_{m} \, \right] &=& \sigma_{m}
\; , \;
\frac{2}{\sqrt{\bar{\Box}}}\left[ \, \epsilon^{ij}\,\partial_{j}E_{mz}+\partial_{t}B_{m}^{\,\,i} \, \right]=-K_{m}^{\,\,i}
\; , \;
(\bar{\nabla}\times{\bf B}_{m})_{z}-\partial_{t}E_{mz}=0
\; .
\;\;\;\;\;\;
\label{eqsM}
\end{eqnarray}
\end{subequations}
Equations (\ref{eqsE}) and (\ref{eqsM}) are invariant under the duality symmetry $(E_{e}^{\,\,i},B_{ez}) \rightarrow (B_{m}^{\,\,i},-E_{mz})$, $(\sigma_{e},{\bf K}_{e}) \rightarrow (\sigma_{m},{\bf K}_{m})$ and $(B_{m}^{i},E_{mz})\rightarrow (-E_{e}^{\,\,i},B_{ez})$, $(\sigma_{m},{\bf K}_{m}) \rightarrow (-\sigma_{e},-{\bf K}_{e})$.
The gauge symmetry is represented by the transformations of the electric and magnetic potentials, {\it i.e.}, $A^{\prime}_{e\bar{\mu}}=A_{e\bar{\mu}}-\partial_{\bar{\mu}}\Lambda_{e}$ and $A^{\prime}_{m\bar{\mu}}=A_{m\bar{\mu}}-\partial_{\bar{\mu}}\Lambda_{m}$, for any two scalar functions $\Lambda_{e}$ and $\Lambda_{m}$, respectively.   
The total EM field is defined by $F_{\bar\mu\bar\nu}=F_{e\bar\mu\bar\nu}+F_{m\bar\mu\bar\nu}$ , that obviously is gauge invariant.
The variational principle in relation to the potentials $A_{e\bar{\mu}}$ and $A_{m\bar{\mu}}$ (that are treated as independent) 
is insured by the lagrangian density   
\begin{eqnarray}
{\cal L}_{em}=-\frac{1}{4} \, F_{e\bar{\mu}\bar{\nu}}\left[\frac{2}{\sqrt{\bar{\Box}}}F_{e}^{\;\;\bar{\mu}\bar{\nu}}\right]
-\frac{1}{4} \, F_{m\bar{\mu}\bar{\nu}}\left[\frac{2}{\sqrt{\bar{\Box}}}F_{m}^{\;\;\,\bar{\mu}\bar{\nu}}\right]
-j_{e\bar{\mu}}\,A_{e}^{\;\,\,\bar{\mu}}
-j_{m\bar{\mu}}\,A_{m}^{\;\,\,\bar{\mu}} \; ,
\end{eqnarray}
whose the correspondent action in $1+2$ dimensions leads to the field equations (\ref{EqsFe}) and (\ref{EqsFm}).
Considering just the magnetic monopole at rest, 
the equations for the magnetism are reduced to
%
\begin{eqnarray}
\frac{2}{\sqrt{-\bar{\nabla}^2}}\left[ \, \bar{\nabla}\cdot{\bf B}_{m} \, \right]=\sigma_{m}
\quad , \quad
(\bar{\nabla}\times{\bf B}_{m})_{z}=0
\; ,
\label{eqsMestatic}
\end{eqnarray}
%
in which the ${\bf B}_{m}$-field is related with the scalar magnetic potential as ${\bf B}_{m}=-\bar{\nabla}\Phi_{m}$, that satisfies the non-local Poisson equation 
\begin{eqnarray}\label{eqPhim}
\frac{2}{\sqrt{-\bar{\nabla}^2}}\left[ \, \bar{\nabla}^2\,{\Phi}_{m} \, \right]=-\sigma_{m} \; .
\end{eqnarray}
The solution of (\ref{eqPhim}) is given by
\begin{eqnarray}
\Phi_{m}({\bf r})=\frac{1}{4\pi} \, \int_{\mathcal{S}}d^2{\bf r}^{\prime} \, \frac{\sigma_{m}({\bf r}^{\prime})}{|{\bf r}-{\bf r}^{\prime}|} \; ,
\end{eqnarray}
that leads to the usual result for the magnetic field of a like point magnetic charge $q_m$ at rest, {\it i.e.}, 
with $\sigma_{m}({\bf r})=q_m\,\delta^{2}({\bf r})$, the magnetic potential is reduced to $\Phi_{m}(r)=q_m/(4\pi r)$, 
whose correspondent field is 
\begin{eqnarray}
{\bf B}_{m}=\frac{q_{m}\,\hat{{\bf r}}}{4\pi \, r^2} \; ,
\end{eqnarray}
The solutions for the dynamical fields are obtained from the electric and magnetic potentials. In terms of $A_{e\bar\mu}$ and $A_{m\bar\mu}$, the equations with sources in the Lorenz gauge $\partial_{\bar\mu}A_{e}^{\,\,\bar\mu}=\partial_{\bar\mu}A_{m}^{\,\,\bar\mu}=0$ are written as
%
\begin{eqnarray}
2\sqrt{\bar{\Box}}\,A_{e}^{\bar\nu}=j_{e}^{\,\,\bar\nu}
\hspace{0.5cm} , \hspace{0.5cm}
2\,\sqrt{\bar{\Box}}\,A_{m}^{\,\,\bar\nu}=j_{m}^{\,\,\bar\nu} \; ,
\end{eqnarray}
%
whose the retarded solutions are, respectively, 
\begin{eqnarray}
A_{e}^{\,\,\bar\mu}({\bf r},t)=\frac{1}{4\pi}\int_{{\cal S}} d^2{\bf r}^{\prime} \, \frac{j_{e}^{\bar\mu}({\bf r}^{\prime},t_r)}{|{\bf r}-{\bf r}^{\prime}|}
\hspace{0.5cm} , \hspace{0.5cm}
A_{m}^{\,\,\bar\mu}({\bf r},t)=\frac{1}{4\pi}\int_{{\cal S}} d^2{\bf r}^{\prime} \, \frac{j_{m}^{\bar\mu}({\bf r}^{\prime},t_r)}{|{\bf r}-{\bf r}^{\prime}|} \; ,
\end{eqnarray}
in which $t_r=t-|{\bf r}-{\bf r}^{\prime}|$ is the retarded time, as defined previously.

The conservation laws can be obtained from the field equations (\ref{eqsE}) and (\ref{eqsM}), using the similar approach from the section (\ref{sec3}). Multiplying the source equations by $F_{e\bar\nu\bar\rho}$ and $F_{m\bar\nu\bar\rho}$, respectively, the sum of these equations with the Bianchi identities yields the continuity equation in the presence of external sources 
\begin{eqnarray}\label{EqTemJ}
\partial^{\bar\mu}T_{em\bar\mu\bar\rho}=j_{e}^{\,\,\bar\nu}
\;F_{e\bar\nu\bar\rho}+j_{m}^{\,\,\bar\nu}\;F_{m\bar\nu\bar\rho} \; ,  
\end{eqnarray}
where the energy-momentum tensor $T_{em\bar\mu\bar\rho}$ contains the strength field tensors defined previously
\begin{eqnarray}
T_{em}^{\;\;\,\,\bar\mu\bar\nu}&=&\bigg[\,\frac{2}{\sqrt{\bar{\Box}}}F_{e}^{\;\,\,\bar\mu\bar\alpha}\,\bigg]F^{\,\,\bar\nu}_{e\;\;\bar\alpha} 
+\bigg[\,\frac{2}{\sqrt{\bar{\Box}}}F_{m}^{\;\,\,\bar\mu\bar\alpha}\,\bigg]F^{\,\,\bar\nu}_{m\;\;\bar\alpha} 
\nonumber \\
&&
+ \frac{1}{4} \;\eta^{\bar\mu\bar\nu} \, \left\{ \bigg[ \, \frac{2}{\sqrt{\bar{\Box}}} F_{e}^{\;\,\,\bar\alpha\bar\beta} \, \bigg] F_{e\bar\alpha\bar\beta}
+\bigg[ \, \frac{2}{\sqrt{\bar{\Box}}} F_{m}^{\;\,\,\bar\alpha\bar\beta} \, \bigg] F_{m\bar\alpha\bar\beta} \, \right\} \; .
\end{eqnarray}
The spatial component $\bar\rho=j$ in (\ref{EqTemJ}) yields the Lorentz force density
\begin{eqnarray}
f_{em}^{i}=\sigma_{e} \, E_{e}^{\,\,i}+\varepsilon^{ij}\,j_{e}^{\,j}\,B_{ez}
+\sigma_{m}\,B_{m}^{\,\,i}-\varepsilon^{ij}\,j_{m}^{\,j}\,E_{mz} \; .
\end{eqnarray}
In the free space, without sources, the expression (\ref{EqTemJ}) is the conservation law $\partial^{\bar\mu}T_{em\bar\mu\bar\rho}=0$, whose the conserved quantities are read as
\begin{subequations}
\begin{eqnarray}
U_{em} &=& \int_{\mathcal{S}} d^2\bar{x} \; \left[ \, \textbf{E}_{e} \cdot\bigg(\frac{1}{\sqrt{\bar{\Box}}} \textbf{E}_{e} \bigg) + B_{ez} \bigg(\frac{1}{\sqrt{\bar{\Box}}} B_{ez} \bigg)
\right.
\nonumber \\
&&
\left.
+\textbf{B}_{m} \cdot\bigg(\frac{1}{\sqrt{\bar{\Box}}} \textbf{B}_{m} \bigg) 
+ E_{mz} \bigg(\frac{1}{\sqrt{\bar{\Box}}} E_{mz} \bigg)
\, \right] \; , 
\label{Uem}
\\
P_{em}^{i} &=& \int_{\mathcal{S}} d^2\bar{x} \; \left\{ \, \varepsilon^{ij} \, \left[ \, \frac{2}{\sqrt{\bar{\Box}}}E_{e}^{\,\,j} \, \right] B_{ez} 
+ \varepsilon^{ij} \, \left[ \, \frac{2}{\sqrt{\bar{\Box}}}B_{m}^{\,\,j} \, \right] E_{mz} \, \right\} \; .
\label{Poyntingem}
\end{eqnarray}
\end{subequations}
The modified angular momentum is defined by the tensor
\begin{eqnarray}
\tilde{M}_{em}^{\;\;\;\;\,\bar\mu\bar\nu\bar\lambda}=T_{em}^{\;\;\;\;\,\bar\mu\bar\lambda}\,x^{\bar\nu}-T_{em}^{\;\;\;\;\,\bar\mu\bar\nu}\,x^{\bar\lambda} \; ,
\end{eqnarray}
that in the absence of sources satisfies the conservation law $\partial_{\bar\mu}\tilde{M}_{em}^{\;\;\;\;\,\bar\mu\bar\nu\bar\lambda}=0$. 
Thereby, the conserved components are
\begin{eqnarray}
\tilde{N}_{em}^{\;\;\;\,\bar\nu\bar\lambda}=\int_{\cal S} d^2{\bar x} \, \tilde{M}_{em}^{\;\;\;\;\,0\bar\nu\bar\lambda} \, ,
\end{eqnarray}
in which the spatial components $\tilde{N}_{em}^{\;\;\;\,ij}$ define the general angular momentum on ${\cal Z}$-direction as 
$J_{z}^{\,\,em}=\varepsilon^{ij}\,\tilde{N}_{em}^{\;\;\;\,ij}/2$, that in terms of the fields is written as
%
%
\begin{eqnarray}\label{JzEM}
J_{z}^{\,\,em}=-\int_{{\mathcal S}} d^2\bar{x} \, x^{i} \, \left\{ \, \left[ \, \frac{2}{\sqrt{\bar{\Box}}}E_{e}^{\,\,i} \, \right] B_{ez} 
+ \left[ \, \frac{2}{\sqrt{\bar{\Box}}}B_{m}^{\,\, i} \, \right] E_{mz} \, \right\}  \; .
\end{eqnarray}
This expression contains the result (\ref{Jz}) if we set $B_{m}^{\,i}=E_{mz}=0$. Thus, the structure of the PED with magnetic monopoles is constructed through the new gauge potential that contains all the information of charge/current for magnetic distributions, and in principle, are independent on the equations of the electric sector. 
The result (\ref{JzEM}) can lead to a new quantization rule for the magnetic monopole in the pseudo-electrodynamics, that will be investigated in a forthcoming project.

\section{Conclusions}
\label{sec6}

The classical pseudo-electrodynamics (PED) is a non-local electrodynamics that emerges from the dimensional reduction of the Maxwell ED, when the classical sources are constrained on a spatial plane. Consequently, this non-local ED lives in a $1+2$ space-time, and the correspondent gauge field is the candidate to mediate interactions between fermions in bidimensional materials. Thus, many applications in condensed matter physics can be investigated taking this theory as the electromagnetic sector in weyl semimetals, graphene, and others. We examine the classical aspects of the pseudo-electrodynamics (PED), as the conservation laws in the complete form, and also interesting results of the propagation fields in the radiation zone. We apply the infinitesimal symmetries in the action to obtain the Noether current of the theory. As we expect, the Noether current is not gauge invariant, and posteriorly, we introduce the technics of the Belifante-Rosenfeld tensor. Thereby, the final result is the non-local energy-momentum tensor, that is symmetric and gauge invariant. From this result, the ${\cal Z}$-component for the general angular momentum is calculated.  
From the solutions for the retarded potentials, and the Jefimenko's equations, we explore the solutions of the EM fields for an electric dipole variable with the time in the radiation zone. We calculate the radiation power of an electric dipole with constant acceleration in which the variation of the power over the radial direction is proportional to squared root of acceleration times the radial distance. This result does not reproduce the Larmor's formula due to non-locality of the Poynting vector in the PED.
For end, we show a way to introduce charge and current magnetic distributions in the PED. In $1+2$ dimensions, we include a new set of field equations through the field strength tensor associated with the magnetic current density. For a like point magnetic charge at rest, the equations reproduce the known result of the magnetic field, {\it i.e.}, ${\bf B}=\hat{{\bf r}}\,q_{m}/(4\pi r^2)$, preserving the static result in the PED. We also obtain that the conservation laws are extended in this case to include the contributions of the new fields. As perspective, all these results corroborate to the study of the quantization of the magnetic monopole in PED and it applications in material physics are an interesting investigation in a forthcoming project.


\par
\par
\par
%

\vspace{0.5cm}

{\bf Acknowledgments}:
S. Duque Cesar is grateful to the National Conselho Nacional de Desenvolvimento Científico e Tecnológico (CNPq) for supporting during his Scientific Initiation.


\end{document}